\documentstyle[prb,aps,epsf,floats,twoside]{revtex}  
\begin{document}

\twocolumn[\hsize\textwidth\columnwidth\hsize\csname
@twocolumnfalse\endcsname

\draft

\title{Magnetic-field-enhanced outgoing excitonic 
       resonance in multi-phonon Raman
       scattering from polar semiconductors}

\author{I. G. Lang and A. V. Prokhorov}
\address{A. F. Ioffe Physico-Technical Institute, Russian
Academy of Sciences, 194021 St. Petersburg, Russia}

\author{M. Cardona}
\address{Max Planck Institut f\"ur Festk\"orperforschung,
Heisenbergstrasse 1, D-70569 Stuttgart, Germany}
\author{V. I. Belitsky, A. Cantarero, and S.T.
Pavlov\cite{Pav}}
\address{Departamento de F\'{\i}sica Aplicada, Universidad de
Valencia, Burjasot, E-46100 Valencia, Spain}

\date{\today}

\maketitle

\begin{abstract}
A combined scattering mechanism involving the states of free
electron-hole pairs (exciton continuum) 
and discrete excitons as intermediate states
in the multi-phonon Raman scattering leads to (1) a strong
increase of the scattering efficiency in the presence of a high magnetic
field and to (2) an outgoing excitonic resonance: the two features
are not compatible when only free pairs (leading to a 
strong increase of the 
scattering efficiency under the applied magnetic field) or discrete 
excitons (resulting in the outgoing resonance at the excitonic gap) 
are taken into account.  

\end{abstract}

\vskip 2pc ]

\narrowtext

\section{Introduction}
In a recent publication,\cite{0} we have shown that the strong
outgoing resonance observed 
in high order multi-phonon resonant Raman scattering
(MPRRS) from polar semiconductors can be explained when  high
energy intermediate electronic states belong to the excitonic
continuum (approximated by free electron-hole 
pairs, EHP) and only couple to the bound
excitonic state at the last stage of a scattering process. 
The high probability of decay
into the continuum strongly opposes the MPRRS mechanism involving
discrete excitons as the only intermediate states for explanation of   
the observed outgoing resonance at the ground
excitonic transition (see Ref.~\onlinecite{0} and references
therein). Cooled by the emission of a sufficiently large number of  
LO-phonons, the EHP binds into an exciton whose energy is not 
enough for LO-phonon-assisted decay.

In this work we analyze the effects of a high magnetic field
on the outgoing excitonic resonance considering, as in
Ref.~\onlinecite{0}, the monomolecular creation of 
a cold exciton by the light-generated free EHPs 
which lose energy but preserve their spatial 
correlation through  the interaction with LO-phonons.   

\section{Model}
\begin{figure}[t]
\vspace{-3cm}
\epsfxsize=3 in
\epsffile{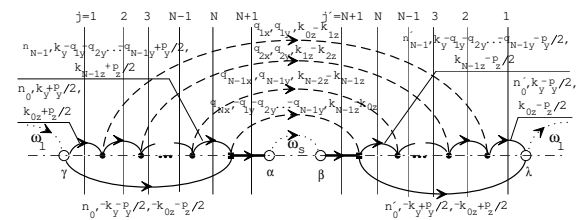}
\vspace{-3cm}
\caption{The diagram involving one discrete exciton
intermediate state contributing to the MPRRS  
efficiency in the range of outgoing resonance. 
Hollow circles represent photon-electron-hole pair interaction, 
bold circles
correspond to the electron-LO-phonon interaction while the
square vertices are shown for discrete-continuum transitions. Solid
lines above (below) the dash-dotted line represent the electrons
(holes) and horizontal lines stay for bound excitons. Bold
dashed lines connected left and right hand sides of the
diagrams correspond to LO-phonons while the dotted lines
represent 
incident (on the left and right sides) and scattered (in the
center) photons. 
\label{1f}}
\end{figure}

\begin{figure}[t]
\vspace{-3cm}
\epsfxsize=3 in
\epsffile{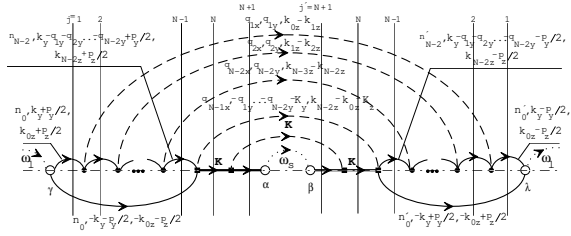}
\vspace{-3cm}
\caption{The diagram with two discrete exciton
intermediate states contributing to the MPRRS  
efficiency in the range of outgoing resonance. 
The square vertices are shown for transitions between two states
of discrete exciton and for discrete-continuum transitions. 
\label{2f}}
\end{figure}

The main contribution to the $N$-th order MPRRS 
efficiency follows from 
processes with one (Fig.~\ref{1f}) and two (Fig.~\ref{2f}) 
bound excitonic intermediate 
states at the last stage of the elementary scattering process. 
Only these contributions correspond to
the cascade of transitions where the bound exciton 
cannot decay into the EHP continuum through the emission of 
LO-phonons. We assume $m_h>>m_e$ so that
the hole energy is less than the energy of one LO-phonon and all 
phonons emitted by the EHP before its 
binding into a discrete exciton are
emitted by the electron.  
 
We use the Landau gauge ${\bf A}={\bf A}(0, xH, 0)$ for a magnetic
field directed along the $z$-axis and the corresponding wave functions
of free EHPs. Only the ground state of the bound exciton is
taken into account for discrete excitonic intermediate states in the
last stage of the process.       
According to Refs.~\onlinecite{1} and \onlinecite{2}, 
the bound exciton wave function in a high magnetic field 
($a>>a_H$, where $a$ is Bohr radius in a zero
magnetic field and $a_H$ is the magnetic length,
$a_H=\sqrt{\hbar c/eH}$) can be written as   
$$
\Psi^{exc}_{{\bf K}_{\perp}K_z}=\Psi_{\perp {\bf K}_{\perp}}
\Psi_{\parallel K_z}~, 
$$
where 
\begin{eqnarray} 
\label{2}
\Psi_{\perp {\bf K}_{\perp}}&=&
{\exp{\left[-\left|{\bf r}_{\perp}-{\bf r}_{\perp}({\bf
K}_{\perp})\right|^2/(4a_H^2)\right]}\over a_H\sqrt{2\pi L_xL_y}}\nonumber\\
&\times&\exp\left\{i\left[(K_x-(y/a_H^2))R_x+K_yR_y\right.\right.\nonumber\\
&+&\left.\left.\Phi({\bf r}_{\perp},
-{\bf K}_{\perp})+C(K_x, K_y)\right]\right\}~,
\end{eqnarray}
and 
\begin{eqnarray}
\Phi({\bf r}_{\perp},-{\bf K}_{\perp})&=&
\left(xy/a_H^2-K_xx-K_yy\right)(m_e-m_h)/2M~,\nonumber\\
C(K_x, K_y)&=&a_H^2K_xK_y(m_e-m_h)/2M~,
\end{eqnarray}
$m_e$ $(m_h)$ is the electron (hole) effective mass,
$M=m_e+m_h$, 
${\bf r}_{\perp}({\bf K}_{\perp})=(a_H^2/H)\left[{\bf
H}\times{\bf K}_{\perp}\right]$   
and ${\bf R}$, ${\bf r}$ are the
center of mass and relative motion coordinates of an electron and hole. 
The longitudinal part $\Psi_{\parallel K_z}$ of the exciton wave
function can be written as 
\begin{equation}
\label{5}
\Psi_{\parallel K_z}={1\over\zeta\sqrt{a_{\parallel}L_z}}
\exp(iK_zR_z)\xi (z/a_{\parallel})~,
\end{equation}
where $\xi (s)$ describes the relative motion of the electron
and hole along the magnetic field direction 
and satisfies the equation $\xi (s=0)=1.$ 
The constant $\zeta$ is determined by the normalization 
condition $$\zeta^2=\int_{-\infty}^{\infty}ds~\xi^2(s)~.$$ 
We do not specify the exact 
form of $\xi (s)$ (see Ref.~\onlinecite{1}) and introduce
the two functions $$\Theta (\alpha)=\int_{-\infty}^{+\infty}ds\exp{(i\alpha
s)}\xi (s)$$ and $$\eta 
(\alpha)=\int_{-\infty}^{+\infty}ds\exp{(i\alpha s)}\xi^2(s)$$ to
be used below. For $\xi (s)=\exp{(-|s|)}$ one finds $\zeta =1$,
$\eta (\alpha)=1/(1+\alpha^2/4)$ and $\Theta
(\alpha)=2/(1+\alpha^2)$.   
The wave function of Eq.~(\ref{2}) reduces to that 
of Ref.~\onlinecite{1} when the Landau gauge is changed for
the symmetric one.   
\section{Scattering efficiency}
The scattering efficiency can be written as\cite{base}  
\begin{equation}
\label{122}
{d^{2}S\over d\Omega d\omega_{s}}={\omega_{s}^{3}\omega_{l}\over
c^{4}}{n_{s}\over n_{l}}e_{s\alpha}^{\ast}e_{s\beta}
e_{l\gamma}e_{l\lambda}^{\ast}S_{\alpha\gamma\beta\lambda}
(\omega_{l},\omega_{s},{\bf\kappa}_l,{\bf\kappa}_s)~,
\end{equation}
where $S_{\alpha\gamma\beta\lambda}$ is the light scattering
tensor of rank four, $c$ the light
velocity in vacuum, $n_{l}$ $(n_{s})$, ${\bf e}_l$ $({\bf
e}_s)$, 
$\kappa_l$ $(\kappa_s)$, $u_l$ $(u_s)$ are the refractive index, 
polarization vector, wave vector, and group velocity 
of the incident
(scattered) light, respectively. 
Using diagrammatic techniques, similar to those of
Refs.~\onlinecite{0}, \onlinecite{2} and \onlinecite{3}, 
we find for the contributions 
of the diagrams in Fig.~\ref{1f} (Fig.~\ref{2f})  
\begin{eqnarray}
\label{12}
{d^{2}S_{N a(b)}\over d\Omega d\omega_{s}}&=&\sigma_0
{\omega_{s}n_s\over\omega_{l}n_l}
{|{\bf e}_l{\bf p}_{cv}|^2|{\bf e}_s{\bf p}_{cv}|^2\over 
\pi a_H^2m_0^2\hbar^2}\nonumber\\
&\times&{\delta(\omega_l-\omega_s-N\omega_{LO})\over 
(\omega_s-\omega_{1H})^2+(\gamma_{exc H}(0)/2)^2}
{1\over L_{Na(b)}}~,
\end{eqnarray}
where $\sigma_0=(e^2/m_0c^2)^2$ and $\gamma_{excH}(0)$ is the
inverse life time (broadening) of the exciton at the ground state 
with energy 
$E_{1H}=\hbar\omega_{1H}$. According to the
assumption $m_h>>m_e$, the energy and the broadening of the hole
have been neglected in all energy denominators. 
The quantity $L_{Na(b)}$
has dimensions of length and, for 
the diagram in Fig.~\ref{1f}    
\begin{equation}
\label{12a}
{1\over L_{Na}}=\sum_\beta{D_{N\beta}\over
\Lambda_{N\beta}Y_{N\beta}}\Xi_{n_0,
n_{N-1}}~,
\end{equation}
where the index $\beta$ designates the sequence of transitions
made by the electron 
through Landau bands emitting successively
$N-1$ phonons. It represents the set of $n_0,
n_1,\dots,n_{N-1}$ Landau numbers and indices $i_1,
i_2,\dots,i_{N-1}$. Each index $i$ may be $0$ or $1$:
it is zero when the electron does not change the direction of
motion along the magnetic field after the phonon emission and one
when the sign of the velocity is opposite in the states before and
after the phonon emission. In Eq.~(\ref{12a}), $D_{N\beta}$ represents the
integral      
\begin{eqnarray}
\label{14}
D_{N\beta}&=&{1\over K_0l}\int_0^{\infty}
\prod_{j=1}^{N-1}\left[{dx_j\over K_jl}
B_{n_{j-1}n_j}(x_j)\chi^{i_j}(K_{j-1}, K_j, x_j)\right]\nonumber\\ 
&\times&\left\langle B_{n_{N-1}n_0}(x_N)\chi^{i_N}(K_{N-1}, 
K_0, x_N)\right\rangle~, 
\end{eqnarray}
where 
\begin{eqnarray}
\label{15}
K_j&=&\sqrt{2m_e\left(\omega_l-\omega_{gH}-n_j\omega_{eH}
-j\omega_{LO}\right)/\hbar}~,\nonumber\\
l&=&\sqrt{\hbar/(2m_e\omega_{LO})}~,
\end{eqnarray}
\begin{equation}
\label{16}
B_{nn^{\prime}}(x)={\min(n!,n^{\prime}!)\over \max(n!,n^{\prime}!)}
e^{-x}x^{|n-n^{\prime}|}\left[L^{|n-n^{\prime}|}_
{min(n,n^{\prime})}(x)\right]^2~,
\end{equation}
and 
\begin{equation}
\label{17}
\chi^i(K, K^{\prime}, x)=\left[x+a^2_H(K\mp 
K^{\prime})^2/2\right]^{-1}~, 
\end{equation}
$\hbar\omega_{gH}=E_g+\hbar eH/(2\mu c)$, where $\mu=m_em_h/M$ and
$E_g$ is the gap. In Eq.~(\ref{17}), the $-~(+)$ sign corresponds to $i=0$
($i=1$).   
  
Note that $i_N=0$ 
when the direction of motion along magnetic field after emission
of $N-1$ phonons coincides with an initial direction. In this
case, $s=\sum_{n=1}^{N-1}i_n$ is an even number. For odd $s$ the direction
of motion is opposite and $i_N=1$. The symbol 
$\langle\dots\rangle$ corresponds to the average over the directions
of wavevectors ${\bf q}_{1\perp}, {\bf q}_{2\perp},\ldots,{\bf
q}_{N-1\perp}$ in the $xy$-plane, when   
$x_j=a_H^2q_{j\perp}^2/2$ and 
${\bf q}_{N\perp}=-\sum_{i=1}^{N-1}{\bf q}_{i\perp}$. 
We used also 
\begin{equation}
\label{19}
Y_{N\beta}=\prod_{j=0}^{N-1}(2\gamma_j/\alpha\omega_{LO})~,~~
\gamma_j=\gamma_e(n_j, K_j)~, 
\end{equation}
where $\alpha$ is the Fr\"ohlich coupling constant. 
When $\gamma_e(n, |k_z|)$ is determined by the interaction with
LO-phonons, we find 
\begin{eqnarray}
\label{20}
\gamma_e(n, |k_z|)&=&\alpha\omega_{LO}\sum_{n^{\prime}}
(2l|k_z^{\prime}|)^{-1}\int_0^{\infty}dx
B_{nn^{\prime}}(x)\nonumber\\
&\times&
\left[\chi^0(|k_z|, |k_z^{\prime}|, x)+\chi^1(|k_z|, 
|k_z^{\prime}|, x)\right]~,\nonumber\\
k_z^{\prime}&=&
\sqrt{k_z^2+2m_e\left[\omega_{eH}(n-n^{\prime})
-\omega_{LO}\right]/\hbar}~~.
\end{eqnarray}
The sum over $n^{\prime}$ is limited by the condition that 
$k_z^{\prime}$ has to be real. When all $\gamma_0, \gamma_1,
\ldots, \gamma_{N-1}$ are determined by the probability to 
emit an LO-phonon in a real transition, the substitution of
Eq.~(\ref{20}) in Eq.~(\ref{19}) and multiplying the result by
$l^N$ leads to $Y_{N\beta}$ defined
in Eq.~(134) of Ref.~\onlinecite{3}. 
However, close to the resonance $\omega_s=\omega_{1H}$, the
electron emitted $N-1$ LO-phonons occupies the state with the
energy less than the energy of an LO-phonon. In this case 
$\gamma_{N-1}$ is determined by some other weaker
scattering mechanism:  
\begin{equation}
\label{20b}
\gamma_{N-1}=\Gamma_{N-1}~,~~~
\Gamma_{N-1}\ll\gamma_j~,~~~j=0,1,\ldots ,N-2~~. 
\end{equation}
This leads to the result: 
\begin{equation}
\label{21}
Y_{N\beta}=Y_{N-1\beta}{2\Gamma_{N-1}\over\alpha\omega_{LO}}~~. 
\end{equation}
The length $\Lambda_{N\beta}$ is defined as
\begin{equation}
\label{21a}
\Lambda_{N\beta}=f_{N\beta}^{-1}(z=0)~,~~~f_{N\beta}(z)=
f^{i_1i_2\ldots i_{N-1}}(z)~. 
\end{equation}
For example, in the case $N=4$ 
\begin{eqnarray}
\label{22}
f^{010}(z)&=&\left[f^{++--}(z)+f^{--++}
(z)\right]/2~~,\nonumber\\
f^{++--}(z)&=&{1\over\lambda_3}
\int_{-\infty}^{\infty}\left[\prod_{j=0}^{j=2}{dz_j\over\lambda_j}\right]
\Upsilon^+\left({z_0\over\lambda_0}\right)\Upsilon^
+\left({z_1-z_0\over\lambda_1}\right)\nonumber\\
&\times&\Upsilon^
-\left({z_2-z_1\over\lambda_2}\right)
\Upsilon^-\left({z-z_2\over\lambda_3}\right)~,
\end{eqnarray}
\begin{equation}
\label{24}
\Upsilon^+(t)=\left\{\begin{array}{cc}
e^{-t},\hspace{0.cm} &t>0\quad; \\
0,\hspace{0.cm} &  t<0\quad, \end{array}\right.~~
\Upsilon^-(t)=\left\{\begin{array}{cc}
0,\hspace{0.cm} &t>0\quad; \\
e^t,\hspace{0.cm} &  t<0\quad, \end{array}\right.
\end{equation}
$\lambda_j=\hbar K_j/m_e\gamma_j$ and $\gamma_j=\gamma_e(n_j,
K_j)$. 

In $p$-representation $f^{++--}(z)$ can be written as 
\begin{eqnarray}
\label{25}
&&~~~~f^{++--}(z)={1\over
2\pi}\int_{-\infty}^{\infty}dp\exp{(ipz)}\nonumber\\ 
&&\left[(1+i\lambda_0p)(1+i\lambda_1p)(1-i\lambda_2p)
(1-i\lambda_3p)\right]^{-1}~.
\end{eqnarray}
Finally, we have used the definition
\begin{eqnarray}
\label{26}
\Xi_{n_0, n_{N-1}}&=&{1\over\zeta^{4}}[\delta_{n_0, 0}
\Theta^2(a_{\parallel}K_0)+
\delta_{n_{N-1}, 0}\Theta^2(a_{\parallel}K_{N-1})\nonumber\\
&-&2\delta_{n_0, 0}\delta_{n_{N-1}, 0}\Theta(a_{\parallel}K_0)
\Theta(a_{\parallel}K_{N-1})], 
\end{eqnarray}
where $\Theta (\alpha)$ has been defined after Eq.~(\ref{5}).  

We proceed to calculate the contribution of the diagram in
Fig.~\ref{2f}. Since one 
of the intermediate states for the process of Fig.~\ref{2f} 
corresponds to an exciton with ${\bf K}\not= 0$, 
we need to comment on some details of the
ground exciton dispersion $E_{exc}(K_{\perp},|K_z|)$. The energy   
$E_{exc}(K_{\perp},|K_z|)$ can be written as 
\begin{equation}
\label{27}
E_{exc}(K_{\perp}, |K_z|)=E_{gH}+E(K_{\perp})+\hbar^2K_z^2/(2M)~.
\end{equation}
The function $E(K_{\perp})$ in some limits can be found
in Ref.~\onlinecite{1}. For our purposes it suffices to note that
$E(K_{\perp}=0)=-\Delta E_{1H}$ is the exciton binding energy in
a high magnetic field and $E_{1H}=E_{gH}-\Delta E_{1H}$. The 
contribution of the diagram in Fig.~\ref{2f} is given by 
Eq.~(\ref{12}), where  
\begin{eqnarray}
\label{30}
{1\over L_{Nb}}&=&\alpha\omega_{LO}{M\over m_e}\int_0^{x_{max}}dx\exp{(-x)}
\sum_{\beta}{R_{N-1\beta}(x,
K_{z0})\over\Lambda_{N-1\beta}Y_{N-1\beta}}\nonumber\\
&\times&{[\eta(a_{\parallel}K_{z0}m_h/M)-
\eta(a_{\parallel}K_{z0}m_e/M)]^2
\over\zeta^4\left(x+a_H^2K_{z0}^2/2\right)^2
lK_{z0}\gamma_{exc}(x, K_{z0})}~~, 
\end{eqnarray}
$K_{z0}=\sqrt{2M\left[\omega_l-\omega_{gH}-E(x)/
\hbar -(N-1)\omega_{LO}\right]/\hbar}$  
is the absolute value of the $z$-component of an exciton wave
vector in the N-th real intermediate state,    
$x=a_H^2K_{\perp}^2/2$, and $\eta(\alpha)$ has been defined
after Eq.~(\ref{5}).   The $K_{\perp 
max}$ is a maximum value of $K_{\perp}$ allowed by energy
conservation, i.e., under the condition that 
$K_{z0}$ is real. At variation of $K_{\perp}$ from zero to infinity the
value $E(x)$ changes from $-\Delta E_{1H}$ to zero.\cite{haha}
Therefore, in the range  
$\omega_l>\omega_{gH}+(N-1)\omega_{LO}$,  
we have $x_{max}\rightarrow\infty$, whereas for 
$\omega_{1H}+(N-1)\omega_{LO}<\omega_l
<\omega_{gH}+(N-1)\omega_{LO}$
the values of $K_{\perp max}$ and $x_{max}$ are determined by
the equation 
$\hbar\omega_l-E_{gH}-(N-1)\hbar\omega_{LO}=E(K_{\perp max})$. 
We used also the following definitions: 

\begin{eqnarray}
\label{36}
&&R_{N-1\beta}(K_{\perp} ,K_{z0})=\nonumber\\
&&{1\over lK_0}\int_0^{\infty}\prod_{j=1}^{N-2}\left[{dx_j\over lK_j}
B_{n_{j-1}n_j}(x_j)\chi^{i_j}(K_{j-1}, K_j, x_j)\right]\nonumber\\
&\times&\left\langle B_{n_{N-2}n_0}(x_{N-1})P_{\beta}
(K_{\perp}, K_{z0})\right\rangle
~~,
\end{eqnarray}
\begin{equation}
\label{37}
P_{\beta}(K_{\perp}, K_{z0})=\left[\Upsilon_{\beta}(K_{\perp}, 
K_{z0})+\Upsilon_{\beta}(K_{\perp}, -K_{z0})\right]/2~~,
\end{equation}
and 
\begin{eqnarray}
\label{38}
&&\Upsilon_{\beta}(K_{\perp}, K_{z0})={1\over\zeta^{4}}\nonumber\\
&\times&\left\{x_{N-1}+(a_H^2/2)\left[(-1)^pK_{N-2}-K_0
-K_{z0}\right]^2\right\}^{-1}\nonumber\\
&\times&\left\{\delta_{n_0,0}\Theta^2\left[a_{
\parallel}\left(K_0+m_hK_{z0}/M\right)\right]
\right.\nonumber\\
&+&
\delta_{n_{N-2},0}\Theta^2\left[a_{\parallel}\left((-1)^pK_{N-2}+
m_eK_{z0}/M\right)\right]\nonumber\\
&-&2\delta_{n_0,0}\delta_{n_{N-2},0}\Theta\left[a_{\parallel}
\left(K_0+{m_h\over M}K_{z0}\right)\right]\nonumber\\
&\times&\left.\Theta\left[a_{\parallel}\left((-1)^pK_{N-2}
-{m_e\over M}K_{z0}\right)\right]\right.\nonumber\\
&\times&\left.\cos{\left[a_H^2\left({\bf q}_{N-1}
\times{\bf K}_{\perp}\right)\right]}\right\}~,
\end{eqnarray}
where $\beta $ is a set of indexes $n_0, n_1,\ldots, n_{N-2},
i_1, i_2,\ldots, i_{N-2}$, $p=\sum_{n=1}^{N-2}i_n$.
The variables of integration in Eq.~(\ref{36}) are 
$x_j=a_H^2q_{j\perp}^2/2$ with a constraint 
${\bf q}_{N-1\perp}=-\sum_{i=1}^{N-2}{\bf q}_{i\perp}
-{\bf K}_{\perp}$. The symbol
$\langle\cdots\rangle$ denotes the average over angles which
determine the direction of vectors ${\bf q}_{1\perp}, {\bf
q}_{2\perp},\ldots, {\bf q}_{N-2\perp}, {\bf K}_{\perp}$ in the
$xy$-plane. 

\section{Applicability limits}

Let us discuss the applicability limits of the expressions
for contributions of diagrams in Fig.~\ref{1f} and 
Fig.~\ref{2f}. We assume that $\Delta
E_H<\hbar\omega_{LO}$ and consider four intervals for the laser
frequency:   
\begin{eqnarray}
\label{40a}
&&(a)~~~~\omega_{1H}+(N-1)\omega_{LO}<\omega_l
<\omega_{gH}+(N-1)\omega_{LO}~,\nonumber\\
&&(b)~~~~\omega_{gH}+(N-1)\omega_{LO}
<\omega_l<\omega_{1H}+N\omega_{LO}~,\nonumber\\
&&(c)~~~~\omega_{1H}+N\omega_{LO}<\omega_l
<\omega_{gH}+N\omega_{LO}~,\nonumber\\
&&(d)~~~~\omega_l>\omega_{gH}+N\omega_{LO}~. 
\end{eqnarray}
The width of the intervals (a) and (c) is   
$\Delta E_H/\hbar$ and the one of interval 
(b) is $(\omega_{LO}-\Delta E_H/\hbar)$. 
The outgoing excitonic resonance coincides with 
the border of intervals (b)
and (c). 

Equations (\ref{12}) and (\ref{12a}) for the contribution of
the diagram in Fig.~\ref{1f} are valid when $K_{N-1}$ 
is real for $n_{N-1}=0$ (see
Eq.~(\ref{15})). This is satisfied within the intervals 
(b), (c) and (d).  
In intervals (b) and (c), the value of $K_N(n_N=0)$ is pure imaginary  and
it is real in interval (d). This means that $\gamma_{N-1}$ is determined
by Eq.~(\ref{20b}) in (b) and (c) (therefore, in the vicinity of
the outgoing resonance) and Eq.~(\ref{21}) is valid.
Let us show that the $\Gamma_{N-1}$ cancels out of the expression
for the contribution of the diagram in Fig.~\ref{1f}. To do this note that
$\Lambda_{N\beta}$ from Eq.~(\ref{21a}) is proportional to the
mean free path   
\begin{equation}
\label{41}
{\cal L}_{N-1}={\hbar K_{N-1}\over m_e\Gamma_{N-1}}~,
\end{equation}
when ${\cal L}_{N-1}>>\lambda_j,~j=0,1,\ldots,N-2$~.  
Thus,  
\begin{equation}
\label{42}
\Lambda_{N\beta}^{-1}={\cal L}_{N-1}^{-1}T_{N\beta}~~, 
\end{equation}
where $T_{N\beta}$ is a dimensionless function of
$\lambda_0,\lambda_1,\ldots,\lambda_{N-2}$. For $N=2$ we
find\cite{2} that $(\Lambda_{20})^{-1}=f^0(z=0)=0$,
$(\Lambda_{21})^{-1}=f^1(z=0)=1/(\lambda_0+\lambda_1)$. This
leads to $T_{20}=0$, $T_{21}=1$. Likewise, for $N=3$,  
$(\Lambda_{300})^{-1}=f^{00}(z=0)=0$,
$(\Lambda_{310})^{-1}=f^{10}(z=0)
=\lambda_0/((\lambda_0+\lambda_1)(\lambda_0+\lambda_2))$,
$(\Lambda_{311})^{-1}=f^{11}(z=0)
=\lambda_1/((\lambda_1+\lambda_0)(\lambda_1+\lambda_2))$, 
$(\Lambda_{301})^{-1}=f^{01}(z=0)
=\lambda_2/((\lambda_2+\lambda_0)(\lambda_2+\lambda_1))$. For
$\lambda_2\rightarrow {\cal L}_2$ this lead to $T_{300}=0$, 
$T_{310}=\lambda_0/(\lambda_0+\lambda_1)$,
$T_{311}=\lambda_1/(\lambda_0+\lambda_1)$ and $T_{301}=1$.     
Using Eqs.~(\ref{21}), (\ref{42}) and (\ref{41}) we obtain  
\begin{equation}
\label{44}
(Y_{N\beta}\Lambda_{N\beta})^{-1}=(Y_{N-1\beta})^{-1}
{\alpha\omega_{LO}m_e\over 2\hbar K_{N-1}}T_{N\beta}~~,
\end{equation}
Thus, the quantity $\Gamma_{N-1}$ does not appear in the final
result. 

Equations (\ref{12}) and (\ref{30}) for the contribution of
Fig.~\ref{2f} are valid when $K_{z0}$ is real
which is true for all four frequency intervals.  In (b), (c) and
(d) the broadening $\gamma_{N-2}$ is determined by the
probability to emit an LO-phonon, whereas 
$\gamma_{N-2}\rightarrow\Gamma_{N-2}$ in (a), where 
$\Gamma_{N-2}<<\gamma_j$,
$j=0,1,\ldots,N-3$. Note that Eq.~(\ref{30}) does not content 
$\Gamma_{N-2}$ in the interval (a)  
as it was shown above. The contribution of
Fig.~\ref{2f} depends strongly on the behavior of
$\gamma_{exc}(K_{\perp},K_{z0})$ in the denominator of
Eq.~(\ref{30}). This is the inverse relaxation time of the
exciton in the state with energy 
$E_{exc}(K_{\perp},
K_{z0})=\hbar\omega_l-(N-1)\hbar\omega_{LO}$.  
For  
$E_{exc}(K_{\perp}, K_{z0})>E_{1H}+\hbar\omega_{LO}$ 
the value of $\gamma_{exc}(K_{\perp},K_{z0})$ is determined by
the probability to emit an LO-phonon and is proportional to
$\alpha$. However, for 
$E_{exc}(K_{\perp}, K_{z0})<E_{1H}+\hbar\omega_{LO}$, 
the real emission of one LO-phonon is impossible and
$\gamma_{exc}(K_{\perp},K_{z0})$ is determined by other much
weaker processes, so that  
$\gamma_{exc}(K_{\perp}, K_{z0})\rightarrow
\Gamma_{exc}(K_{\perp}, K_{z0}),$ with 
$\Gamma_{exc}<<\gamma_{exc}$. The change of the scattering
mechanism dominating the broadening takes place at the frequency
corresponding to the outgoing excitonic resonance. Below this
point, the contribution of Fig.~\ref{2f} exceeds strongly the one of
Fig.~\ref{1f}. Note that in this range we have to take into account
other contributions involving processes with acoustic phonons
(see below). 

Finally, the pole approximation (i.e., real transitions) 
for integrals over
$k_{0z},k_{1z},\ldots,k_{N-1z}$ for the contribution of
Fig.~\ref{1f} and over $k_{0z},k_{1z},\ldots,k_{N-2z}$ for
Fig.~\ref{2f} results in the constraints $N\ge 2$ and $N\ge 3$ for
Eq.~(\ref{12a}) and Eq.~(\ref{30}), respectively.  

\section{Discussion and conclusions}

Let us consider the resonant behavior of the MPRRS efficiency
as a function of $H$ and $\omega_l$. We limit ourselves to the
case $N\ge 3$ where both Eq.~(\ref{12a}) and Eq.~(\ref{30}) are
valid. Both contributions increase in the vicinity of $K_0=0$, 
which corresponds to  
$\omega_{lmax,
m_h\rightarrow\infty}\simeq\omega_{gH}+n\omega_{eH}$.  
Taking into account the finite value of $m_h$ leads to an exact
relation 
$\omega_{l max}(n)=\omega_{gH}+eHn/\mu c$.  
This condition corresponds to the creation of EHPs in the
vicinity of the Landau band bottoms. The resonant conditions can
be achieved by changing either $H$ or $\omega_l$. The
maxima in a magnetic field dependence take place at  
\begin{equation}
\label{56}
H_{max}(n)={\mu c\over e}{\omega_l-\omega_g\over n+1/2}~~, 
\end{equation}
being independent on the order $N$ of the scattering process. 
There is an additional resonance\cite{2} for $N=3$ corresponding to the
contribution of Fig.~\ref{2f} at 
$\omega^{\prime}_{lmax,
m_h\rightarrow\infty}\simeq\omega_{gH}+neH/m_ec+\omega_{LO}$.  
This resonance follows from the increase of
$(\Lambda_{21})^{-1}=1/(\lambda_0+\lambda_1)=1/(\hbar
K_0/m_e\gamma_0+\hbar K_1/m_e\gamma_1)$ in 
Eq.~(\ref{30}), when $K_1\rightarrow 0$, because of the divergence
in $\gamma_0$ (see Eq.~(\ref{20})). Note also that contribution
of Fig.~\ref{2f} is equal to zero\cite{2} for 
$\omega_{lmin,m_h\rightarrow\infty}(n,N-1) 
=\omega_{gH}+neH/m_ec+(N-1)\omega_{LO}$.  

Above the outgoing resonance the contributions of Fig.~\ref{1f} and
Fig.~\ref{2f} in the MPRRS efficiency are of the same order of
magnitude. However, as it was mentioned before, below the
resonance the contribution of Fig.~\ref{2f} strongly increases
because of the strong increase in the exciton life time in the
real intermediate state with the energy being too small for
emission of an LO-phonon. In this range, other scattering
processes like the absorption of 
LO-phonons, interaction with acoustic phonons, etc. 
have to be taken into account.  We give now
a qualitative picture of the process including into our
consideration the distribution function of excitons with respect
to $K_{\perp},~|K_z|$. Let us introduce the integral efficiency
for the $N$-th order process as 
\begin{equation}
\label{f0}
S_N=\int\int{d^2S_N\over d\Omega d\omega_s}d\Omega d\omega_s
={1\over u_l}\sum_{\bf\kappa_s}{\bar W_{sN}}~,  
\end{equation}
where $u_l$ is the group velocity of incident light and ${\bar
W_{sN}}$ the normalized probability to emit the scattered light
quantum per unit time.\cite{4}  
Equation (\ref{f0}) differs from 
Eq.~(\ref{12}) only by the absence of
the factor $4\pi\delta (\omega_l-\omega_s-N\omega_{LO})$. 

On the other hand, 
\begin{equation}
\label{f2}
\sum_{\bf\kappa_s}\bar{W}_{sN}=\sum_{K_{\perp}, K_z}P_{exc
N-1}(K_{\perp}, |K_z|)\gamma_l(K_{\perp}, |K_z|)~, 
\end{equation}
where $P_{excN-1}(K_{\perp},|K_z|)$ is the normalized dimensionless 
distribution function of excitons created by the light in a N-1
LO-phonon-assisted process and 
$\gamma_l(K_{\perp},|K_z|)$ the probability of 
an LO-phonon-assisted emission of 
the scattered light quantum which can be 
written as $\gamma_l(K_{\perp}, |K_z|)=\sum_{\bf\kappa_s}w_s$ 
and   
\begin{equation}
\label{f4}
w_s={2\pi\over\hbar}\sum_f\left|\sum_a{\langle f|U_s|a\rangle
\langle a|{\cal H}_{int}|i\rangle\over
E_i-E_a+i\hbar\gamma_a/2}\right|^2\delta(E_i-E_f)~, 
\end{equation}
${\cal H}_{int}$ is Fr\"ohlich interaction of the exciton with
LO-phonons and $U_s$ represents the interaction of excitons with the
light. The initial and final state energy is 
$E_i=E_{exc}(K_{\perp},|K_z|)$ and
$E_f=\hbar\omega_s+\hbar\omega_{LO}$, respectively. The intermediate
state energy $E_a=E_{aEHP}+\hbar\omega_{LO}$ includes both the
discrete and continuum part of the excitonic dispersion. Let us
separate $\gamma_l$ in two corresponding parts,  
$\gamma_l=\gamma_{ldisc}+\gamma_{lcont}$. The outgoing resonance
is related to the contribution $\gamma_{ldisc0}$ 
to $\gamma_{ldisc}$ coming from the 
transition via ground state of the exciton. According to
Eq.~(\ref{f4}), we find  

\begin{eqnarray}
\label{f5}
&&\gamma_{l disc 0}(K_{\perp}, |K_z|)=4{n_s\omega_s\over \hbar c^3}
\left({e\over m_0}\right)^2|{\bf e}_s{\bf p}_{cv}|^2\nonumber\\
&&~~~~~~~{\alpha\omega_{LO}^2l\over
K_{\perp}^2+K_z^2}
{\exp{\left(-a_H^2K^2_{\perp}/2\right)}\over 
a_{\parallel}a_H^2\zeta^6}\nonumber\\
&&{[\eta (K_za_{\parallel}m_h/M)-
\eta (K_za_{\parallel}m_e/M)]^2\over 
[\omega_{exc}(K_{\perp}, |K_z|)-\omega_{1H}-\omega_{LO}]^2+
\gamma^2_{exc H}(0)/4}~. 
\end{eqnarray}
Above the excitonic resonance, $\omega_l>\omega_{1H}
+N\omega_{LO}$, we have 
\begin{equation}
\label{f6}
P_{exc N-1}(K_{\perp}, |K_z|)={W_{exc N-1}(K_{\perp}, |K_z|)\over
\gamma_{exc}(K_{\perp}, |K_z|)}~~, 
\end{equation}
where $W_{exc N-1}(K_{\perp}, |K_z|)$ is the normalized number of excitons
created per unit time in the volume $V_0$ in the $N-1$
LO-phonon-assisted process. The probability $W_{exc N-1}(K_{\perp}, |K_z|)$   
has been calculated in Ref.~\onlinecite{2} for
$N=4$. Being used in Eq.~(\ref{f6}) together with
Eqs.~(\ref{f5}), (\ref{f0}) and (\ref{f2}) it reproduces the
result of Eqs.~(\ref{12}) and (\ref{30}).    

Note that above the outgoing resonance the distribution is not
zero only in very narrow interval of energies\cite{2} since
$W_{excN-1}(K_{\perp},|K_z|)$ is proportional to $\delta
[\omega_l-(N-1)\omega_{LO}-E_{exc}(K_{\perp},|K_z|)/\hbar]$.
However, for $\omega_l$ below the resonance the distribution
becomes smooth. If the most important mechanism in this range is
the interaction with acoustic phonons, one has to take into
account diagrams with external acoustic phonon lines. In the
range (b) (see Eq.~(\ref{40a})) the smoothing of the 
distribution should be weaker than in the range (a). The reason
of this is the kinetic energy of exciton in range (b) which is
larger than the exciton binding energy. Since the probability of scattering
and decay via the interaction with phonons are of the same
order, 
the exciton decays after a few interactions with acoustic
phonons. The decay of an
exciton in the range (a) is suppressed because of its small
energy. In this case, the distribution depends on the probability
of non-radiative recombination. At zero magnetic field, the
distribution of excitons in the range (a) has been considered in
Refs.~\onlinecite{5,6,7} and \onlinecite{8}.  

The smoothness of the distribution leads to $(i)$ the broadening of 
the MPRRS peaks in the range (b) and especially in the 
range (a) and to $(ii)$ the increase of the integral scattering
intensity, since the diagrams with acoustic phonon lines give
additional contributions into the MPRRS efficiency. 

To summarize, we have shown that the outgoing excitonic
resonance has to be strongly enhanced under a high magnetic
field. Above the outgoing resonance, the scattering 
efficiency for $N\ge 3$ may be up to $\alpha ^{-2}$ times stronger
than in a zero magnetic field where the MPRRS efficiency\cite{0} is
proportional to $\alpha ^3$, whereas in a high magnetic field it is
proportional to $\alpha$, as it follows from Eqs.~(\ref{12}),
(\ref{12a}) and (\ref{30}). The crossover from  $\alpha ^3$ to 
$\alpha$ results from the quasi-one-dimensional character of
free EHPs in $N$ (Fig.~\ref{1f}) or $N-1$ (Fig.~\ref{2f}) intermediate
states under a high magnetic field. The enhancement is also
valid for the ranges (a) and (b) below the excitonic resonance, where
one has to calculate the exciton distribution function taking 
into account the interaction with acoustic phonons. To the best
of our knowledge such calculations have yet to be performed. However,
the distribution function is proportional to the creation 
probability of excitons with energy in the interval between $0$
and $\hbar\omega_{LO}$ which is increased by $\alpha ^{-2}$
times in a high magnetic field.\cite{2} Thus, the MPRRS
efficiency also increases below the excitonic resonance. 

The integral efficiency as a function of $\omega_l$  
has to be asymmetric with respect to the 
point $\hbar\omega_l=N\hbar\omega_{LO}+E_{1H}$  
because of the strong increase in
the exciton life time below the resonance and appearance of
additional contributions from the processes with acoustic
phonons.

\acknowledgements
V. I. B. and S. T. P. thank the European Union,  
Ministerio de Educacion y
Ciencia de Espa\~na (DGICYT) and the Russian Fundamental
Investigation Fund (93-02-2362, 950204184A)  
for financial support and the
University of Valencia for its hospitality. This work has been
partially supported by Grant PB93-0687 (DGICYT).


%

\begin{references}
\bibitem{haha} For $a_H^2K_{\perp}^2\simeq a^2/a_H^2$ there is
no bound exciton in a high magnetic field ($a>>a_H$). 
However, in the integral of Eq.~(\ref{30}) only the
range $x\le 1$ is important, a fact which allows us to integrate in the
interval between zero and infinity.
\end{references}
\end{document}